# Construction of Chiral Metamaterial with U-Shaped Resonator Assembly

Xiang Xiong[1], Wei-Hua Sun[1], Yong-Jun Bao[2,1], Ru-Wen Peng[1], Mu Wang[1,*], Cheng Sun[2], Xiang Lu[3], Jun Shao[3], Zhi-Feng Li[3], Nai-Ben Ming[1]

[1]National Laboratory of Solid State Microstructures and Department of Physics, Nanjing University, Nanjing 210093, China
[2]Department of Mechanical Engineering, Northwestern University, Evanston, IL 60208-3111, USA
[3]National Laboratory for Infrared Physics, Shanghai Institute of Technical Physics, Chinese Academy of Sciences, Shanghai 200083, China

Chiral structure can be applied to construct metamaterial with negative refractive index (NRI). In an assembly of double-layered metallic U-shaped resonators with two resonant frequencies $\omega_H$ and $\omega_L$, the effective induced electric and magnetic dipoles, which are contributed by the specific surface current distributions, are collinear at the same frequency. Consequently, for left circularly polarized light, NRI occurs at $\omega_H$, whereas for right circularly polarized light it occurs at $\omega_L$. Our design provides a new example to apply chiral structures to tune electromagnetic properties, and could be enlightening in exploring chiral metamaterials.

*Correspondence author. E-mail: muwang@nju.edu.cn

Interaction of electromagnetic waves with sub-wavelength metallic microstructures has attracted much attention in recent decade because of the new physical properties and effects [1-10]. With specially designed metallic microstructures, it is possible to achieve novel electromagnetic properties that are not naturally available. One of the focused features nowadays is negative refractive index (NRI), which was initially proposed as an academic concept by Veselago [11]. In 1990s, Pendry pointed out that it was possible to construct artificial materials with negative refractive index by assembling an array of subwavelength components that resonates with oscillating electric and magnetic fields [12]. Once the real parts of both permittivity ($\varepsilon_r$) and permeability ($\mu_r$) are simultaneously negative, negative refractive index could be realized. Up to now, metamaterial with negative refractive index has been realized with split rings [13,14], fishnets [3,15], and short wire pairs [16,17].

An alternative way to achieve negative refractive index is to utilize chiral material [18-23], where the chirality suppresses the refractive index of light with one handedness, and increases the refractive index of light with the other handedness. Experimentally, chirality-induced negative refractive index has been demonstrated in cross-wire structure [21], twisted rosettes [22], and interlocked split-ring resonators [24]. For example, a terahertz chiral matematerial has been reported by Zhang et al [20], which is essentially a micro-sized inductor-capacitor (LC) resonance system. In their structure the electric and magnetic dipoles are strongly coupled, and can be simultaneously excited. It is noteworthy that for the structure proposed by Zhang *et al*.

the angle between the directions of effective electric and magnetic dipoles is relatively small, which contributes to an evident chiral behavior. We expect that once the angle between the directions of two dipoles vanishes or equals to 180°, the chiral feature of the structure could be more efficiently utilized.

In this communication we demonstrate that in an assembly of double-layered metallic U-shaped resonators (USRs), the effective induced electric and magnetic dipoles, which are contributed by the specific surface current distributions, are aligned in the same (or the opposite) direction at the same frequency. The resulted strong chiral feature leads to negative refraction for circularly polarized light.

The geometry of double-layer-stacked USRs is schematically shown in Fig. 1(a), where the USR on the upper layer and that on the lower layer is orthogonally rotated. The coordinate frame is so set that the diagonal directions of USRs are defined as *x*- and *y*-axes, respectively. In this coordinate frame, as shown in Fig. 1(a), the opening of the upper layer USR points to 45°, while that of the lower layer USR points to -45°. For normal incidence of electromagnetic wave (*z*-axis incidence), resonant dips occur at lower frequency $\omega_L$ (~390 cm$^{-1}$) and higher frequency $\omega_H$ (~590 cm$^{-1}$), respectively, as shown in Fig. 1(b). The surface current distribution at the frequencies of resonance has been calculated. One may easily find that at $\omega_L$, the electric currents on the upper and lower layers of USRs flow in parallel directions, whereas at $\omega_H$ the surface currents on the two layers are anti-parallel, as illustrated in Fig. 1(c) and Fig. 1(d) respectively. It should be pointed out that the relative direction of the induced electric currents can be either parallel or anti-parallel on two layers of a USR pair, and this feature is independent of the polarization of the incident light.

The induced electric current on the surface of metallic microstructures generates an induced electromagnetic field, which interacts with the incident field and consequently leads to some novel optical properties. The effective induced surface current flowing on each layer of metal structure can be schematically illustrated by the highlighted long arrows, as shown in Fig. 1(c) and Fig. 1(d), respectively. The induced surface currents on the upper and the lower layers of USRs can be projected to *x*-axis and *y*-axis, respectively. At lower resonant frequency $\omega_L$ (Fig. 1(c)), the projected currents along *x*-axis on the upper and the lower layers are anti-parallel. The projected currents on the upper and lower layers along *y*-direction are parallel. It follows that the curl integration along the loop in x-z plane is non-zero, indicating that an induced magnetic filed ***H'*** is established along *y*-direction (Ampere law), or, in other word, an effective magnetic dipole along *y*-direction is induced. Along *y*-direction, in y-z plane, two parallel projected currents can be identified on the upper and the lower layers of USRs, suggesting that there is an induced electric filed ***E'*** along -*y*-direction (Ohm law), or, in other word, an effective electric dipole along -*y*-direction is induced. Similar analysis can be applied to surface current at higher resonance frequency $\omega_H$ (Fig. 1(d)), where an induced magnetic field ***H'*** occurs in -*x*-direction, and an induced electric filed ***E'*** occurs also in -*x*-direction.

It is noteworthy that in Fig. 1 the wave vector of the incident light is along z-direction and the induced magnetic and electric dipoles are collinear along one of the diagonal directions (x- or y-axis) of USR, *i.e.*, the effective induced electric and magnetic dipoles are parallel (or anti-parallel), so their induced electric and magnetic fields are parallel (or anti-parallel), allowing optical activity to occur more efficiently.

Four such optically active USR pairs have been assembled into an unit, as shown in Fig. 2(a), where the upper layer and the lower layer of each USR pair have been rotated for 90º clockwise (or anti-clockwise), respectively. Due to the fourfold rotational symmetry of the unit, the transmission and reflection properties of this structure do not rely on the orientation of the sample with respect to the polarization of incident light. Commercial software based on finite difference time domain (FDTD) method (CST Microwave Studio) is applied to calculate the transmission and reflection coefficients of the array made of the chiral units, and the results are shown in Fig. 2(b). In the calculation, the permittivity of gold in infrared regime is based on the Drude model, $\varepsilon(\omega)=1-\omega_p^2/(\omega^2+i\omega_\tau\omega)$, where $\omega_p$ is the plasma frequency and $\omega_\tau$ is the damping constant. For gold, the characteristic frequencies are taken as $\omega_p=1.37\times10^4$ THz, and $\omega_\tau=40.84$ THz [25]. As shown in Fig. 2(b), two resonance dips occur at 390 cm$^{-1}$ and 590 cm$^{-1}$, respectively, which correspond to the transmission of parallel polarization of input and output light ($|t_{//}|$) and to the transmission of perpendicular polarization of input and output light ($|t_\perp|$). For optically nonactive material, when the incident light is polarized along the principle axis, only transmission of $t_{//}$ could be detected and $t_\perp$ vanishes. However, for optically active material, the chiral behavior helps to rotate the polarization of incident light and convert a portion of energy from one polarization to the other. Hence $t_\perp$ will be detected.

Following the regulation, the transmission and reflection coefficients of the left-handed circularly polarized (LCP) light and the right-handed circularly polarized (RCP) light are expressed as $t_L=t_{//}-it_\perp$, $t_R=t_{//}+it_\perp$, $r_L=r_{//}-ir_\perp$, $r_R=r_{//}+ir_\perp$, respectively. In our structure $r_\perp$ is zero (as shown in Fig. 2(b)), suggesting that the polarization of the reflection light does not change.

The calculated transmission coefficients of LCP and RCP ($t_L$ and $t_R$) are shown in Fig. 2(c). Due to the chirality of the structure, the transmission for LCP and RCP split into two different curves (Fig. 2(c)). Two resonant dips appear at $\omega_L = 390$ cm$^{-1}$ and $\omega_H =590$ cm$^{-1}$ in both $t_L$ and $t_R$, respectively. For the resonance at $\omega_L = 390$ cm$^{-1}$, the dip in $|t_R|$ is much deeper than that in $|t_L|$, suggesting that the resonance for RCP is much stronger than that for LCP. For the resonance at $\omega_H$ = 590 cm$^{-1}$, however, the dip of LCP is much deeper than that of RCP. We define here $\delta$ as the phase difference between $t_\perp$ and $t_\parallel$. The dependence of $\delta$ as a function of wave number is illustrated in Fig. 2(d). It follows that an elliptical polarized light is generated [26] when $\delta \neq n\pi$ (n is an integer). Meanwhile, the end of electric field vector revolves clockwise for $\sin\delta >0$ and count-clockwise for $\sin\delta <0$. Consequently, the outgoing light below and above 460 cm$^{-1}$ (which corresponds to $\sin\delta =0$) have different chirality. This fact also suggests that the resonances at lower and higher frequencies have different chirality. The azimuth angle of the principal polarization axis of the transmitted wave, $\theta$, is defined as $\theta = \frac{1}{2}[\arg(t_R)-\arg(t_L)]$ [21,26]. It represents the change of polarization angle when a linearly polarized light is incident on the USRs assembly. As plotted in Fig. 2(d), at at $\omega_L$ the rotation angle reaches -25° and at $\omega_H$ the rotation angle reaches -72°.

It is known that the refractive indices for LCP and RCP lights can be expressed as [19] $n_{R/L} = \sqrt{\varepsilon\mu} \pm \xi$, where $\varepsilon$ and $\mu$ are the effective permittivity and permeability, respectively; $\xi$ describes the coupling between the electric and the magnetic dipoles along the same direction. Following ref. [22], the effective impedance (Z) and refractive index for LCP and RCP light can be derived from the reflection and transmission coefficient and the other material parameters,

$$\xi=(n_R-n_L)/2,$$

$$\mu=Z(n_R+n_L)/2,$$

$$\varepsilon=(n_R+n_L)/2Z.$$

It is noteworthy that for our USR structures, evident drop of refractive index occurs at

$\omega_H$ for LCP light (Fig. 3(a)), and at $\omega_L$ for RCP light (Fig. 3(b)). Yet the refractive index shows much smaller modification for LCP beam at $\omega_L$ and for RCP at $\omega_H$. This is due to the cancellation of the contributions from the product of permittivity and permeability and that from the chirality. Far away from the resonant frequency, the difference between the refractive indices of RCP and LCP diminishes. The effective permittivity and permeability have been retrieved, as illustrated in Fig. 3(c) and (d). At $\omega_L$, the permittivity reaches negative values, whereas the magnetic resonance is not sufficiently strong to provide a negative permeability; at $\omega_H$ the modification of permittivity and permeability are both sufficiently strong, hence negative values are realized. The chiral term $\xi$ and impedance Z are shown in Fig. 3(e) and (f), respectively, which demonstrate strong chirality of our USRs at both $\omega_L$ and $\omega_H$. Therefore, Fig. 3 demonstrates that the negative refractive index for LCP light around $\omega_H$ arises from two origins: one is from the negative value of both permittivity and permeability, and the other is from the charility of the structure ($\xi >0$). The negative refractive index for RCP light around $\omega_L$, however, originates only from the chirality ($\xi <0$). The efficiency of the negative refractive index is usually characterized by figure of merit (FOM), which is defined as FOM=|Re($n$)/Im($n$)|. For LCP light, we find that the negative refractive index at $\omega_H$ has FOM=2.4, while for RCP light, the negative refractive index at $\omega_L$ has FOM=0.5.

Based on the above calculations, we design a type of chiral metamaterial with the array of units of fourfold rotated USRs. For normal incidence light, on each pair of USRs, parallel and anti-parallel oscillating induced surface currents are excited. Consequently, collinearly aligned effective electric dipole and an effective magnetic dipole are generated. These two dipoles are so strongly coupled at both $\omega_L$ and $\omega_H$ that they are simultaneously excited. It is this feature that leads to the optical activity. It is noteworthy that the effective electric and magnetic dipoles are aligned in the opposite direction at $\omega_L$ and in the same direction at $\omega_H$. This difference results in the different handedness at the lower and the higher frequencies. It also contributes to the negative refractive index for RCP light or LCP light, respectively.

The metallic array of units of fourfold rotated USRs are experimentally fabricated with alignment nesting photolithography. A U-shaped pattern is first defined on the substrate using photoresist. A 100nm-thick gold film is then blanket-deposited on the patterned substrate, covering the areas with photoresist and areas where the photoresist has been removed (U-shaped pattern). With solvent the photoresist was removed, leaving only the gold USRs on silicon substrate. A layer of 600-nm-thick silicon nitride is deposited afterwards as a spacer layer. Thereafter, a layer of photoresist is once again spin-coated and followed by alignment nesting lithography. Then the second layer of gold film is deposited and lift-off procedure is used to remove the photoresist, leaving only the second layer of gold USRs. Meanwhile, each USR on the upper layer located exactly above the one on the lower layer, yet the orientation had being rotated for 90º in a specific way. Consequently an array of specifically arranged chiral USR units is fabricated, as shown in Fig. 4(a). For sake of structural symmetry in optical measurements, a 2μm-thick silicon layer is cap-deposited on the sample surface before finishing the sample fabrication.

The system for optical property measurements is schematically shown in Fig. 4(b). The chiral sample is characterized by a vacuum infrared Fourier-transform spectrometer (Bruker Vertex 70v). In the transmission measurement, the sample was placed between two linear polarizers, and angles $\theta_1$ and $\theta_2$ can be independently adjusted. Three sets of $T_{//}$ and $T_{\perp}$ are measured with different $\theta_1$ and $\theta_2$. As shown in Fig. 4(c), the red solid line corresponds to the scenario $\theta_1=\theta_2=$ -45º; the blue solid line corresponds to the scenario $\theta_1=\theta_2=0°$; the black solid line corresponds to the scenario $\theta_1=\theta_2=45°$; the red dash line corresponds to the scenario $\theta_1= -45, \theta_2=45°$; the black dash line corresponds to the scenario $\theta_1= 45, \theta_2= -45°$; and the blue dash line corresponds to the scenario $\theta_1= 90°, \theta_2=0°$, respectively. One may easily find out that all the $T_{//}$ curves are almost identical and all $T_\perp$ curves are almost identical, suggesting that the transmission is polarization-independent. Dips in $T_{//}$ and peaks in $T_\perp$ at lower frequency (140 cm$^{-1}$) and higher frequency (205 cm$^{-1}$) indicate that at these two resonant frequencies, the USRs have rotated the polarization of light. It should be pointed out that for optical inactive material, only transmission $T_{//}$ could be detected and $T_\perp$ vanishes when the incident light is polarized along the principal axis. For our USR resonators, the chiral feature rotates the polarization of incident light and converts a portion of energy from one polarization to the other, which leads to the dips in $T_{//}$ and the peaks in $T_\perp$. Simulations in Fig. 4(d) demonstrate excellent agreement with the experimental measurements.

    The optical activity of chiral structures provides new approach to realize negative refractive index, and has recently shown promising applications in designing new optical devices that may beat the diffraction limit [6,27], and explore highly sensitive sensors [28,29]. Thus far different types of chiral structures have been constructed [20-22,30-36], and an important goal is to decrease the angle between the effective electric and magnetic dipoles, so to increase the efficiency in realizing negative refractive index. The USR structure reported here demonstrates a unique feature that the excited effective electric and magnetic dipoles are collinear. In this way, the chirality of our USRs can be fully utilized.

In conclusion, we report here a chiral structure made of an assembly of double-layerd USRs, which produces negative refraction for LCP at $\omega_H$ and RCP at $\omega_L$. The chirality originates from the collinear excitation of effective electric and magnetic responses at the same frequency. The experimental data are in consistent with the calculations. Our current structure works in infrared frequency, we also expect that it can be reproducible at other frequencies. We suggest that such a design provides a new example to apply chiral structures in tuning the electromagnetic properties, and could be enlightening in exploring chiral metamaterials.

The authors acknowledge the financial supports from MOST of China (2004CB619005 and 2006CB921804), NSF of China (10625417, 10874068 and 50972057) and Jiangsu Province (BK2008012). The discussions with Prof. Xiang Zhang and Dr. Shuang Zhang are also sincerely acknowledged.

**Figure Captions**

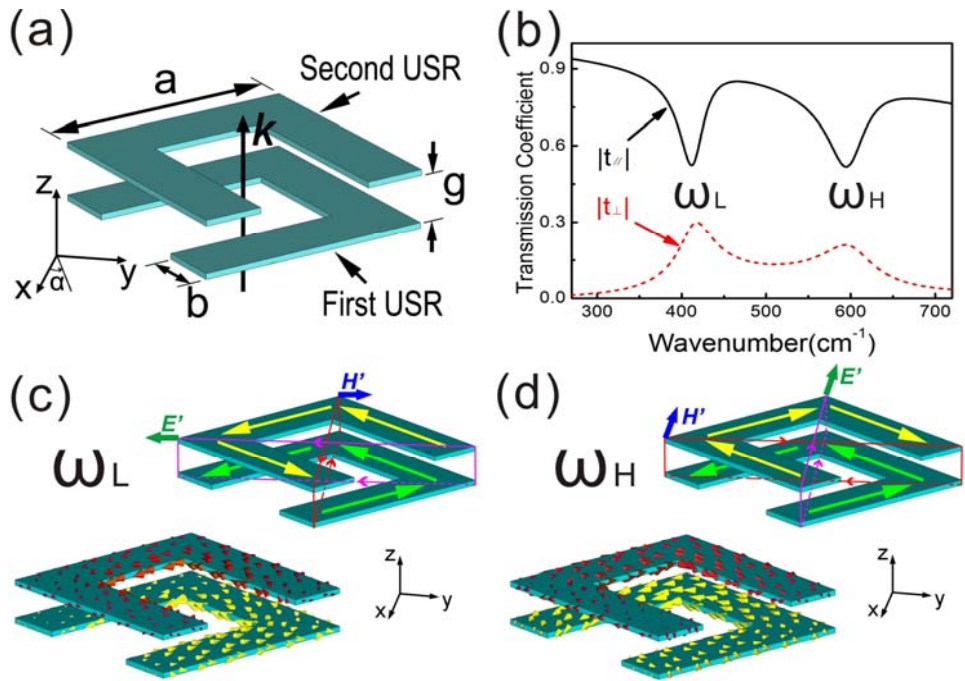

**Fig. 1 (a)** The detail structure of stacked and orthogonally rotated USR pair. a=4.0μm, b=1.0μm, g=0.6μm. **(b)** The transmission coefficient of USRs for normal incidence, and resonance occurs at $\omega_L$ and $\omega_H$. **(c)-(d)** The surface current density excited on USRs at lower ($\omega_L$) and higher ($\omega_H$) resonant frequencies. The small arrows represent the surface current distribution. By setting ***x***- and ***y***-axis along the diagonal directions of USRs, the surface current can be projected along ***x***- and ***y***-directions respectively.

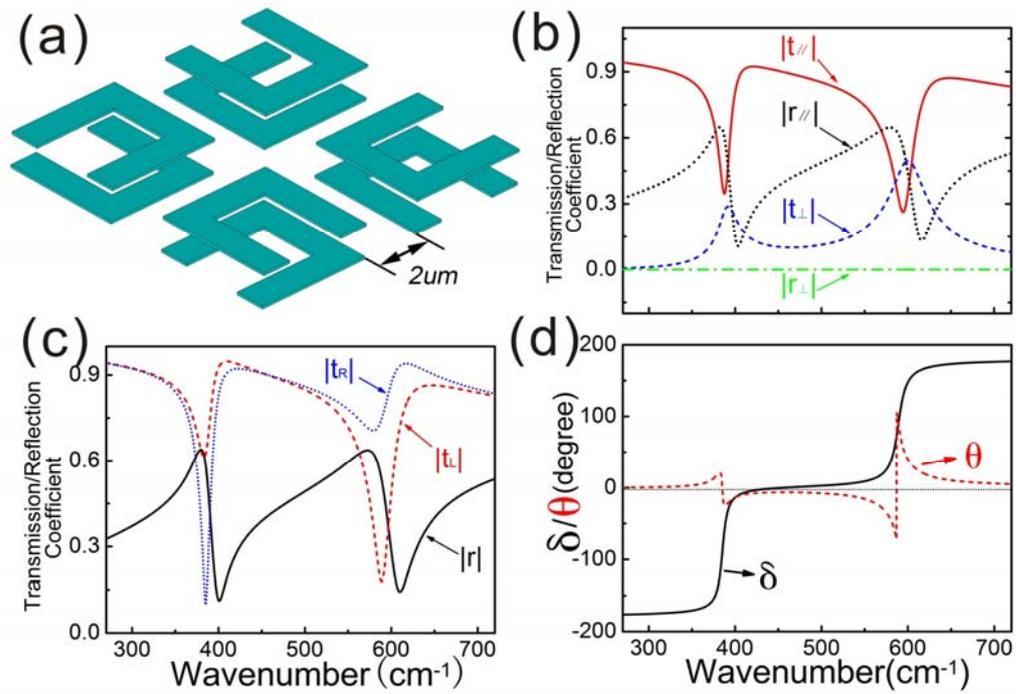

**Fig. 2 (a)** The unit cell constructed by four pairs of USRs. **(b)** The amplitudes of $t_{/\!/}$, $t_\perp$, $r_{/\!/}$ and $r_\perp$. **(c)** The amplitudes of $t_L$, $t_R$ and $r$. **(d)** The plot to show the phase difference $\delta$ between $t_\perp$ and $t_{/\!/}$ as a function of wavenumber, and the azimuth angle of the principal axis of polarization of transmission wave $\theta$ as a function of wavenumber.

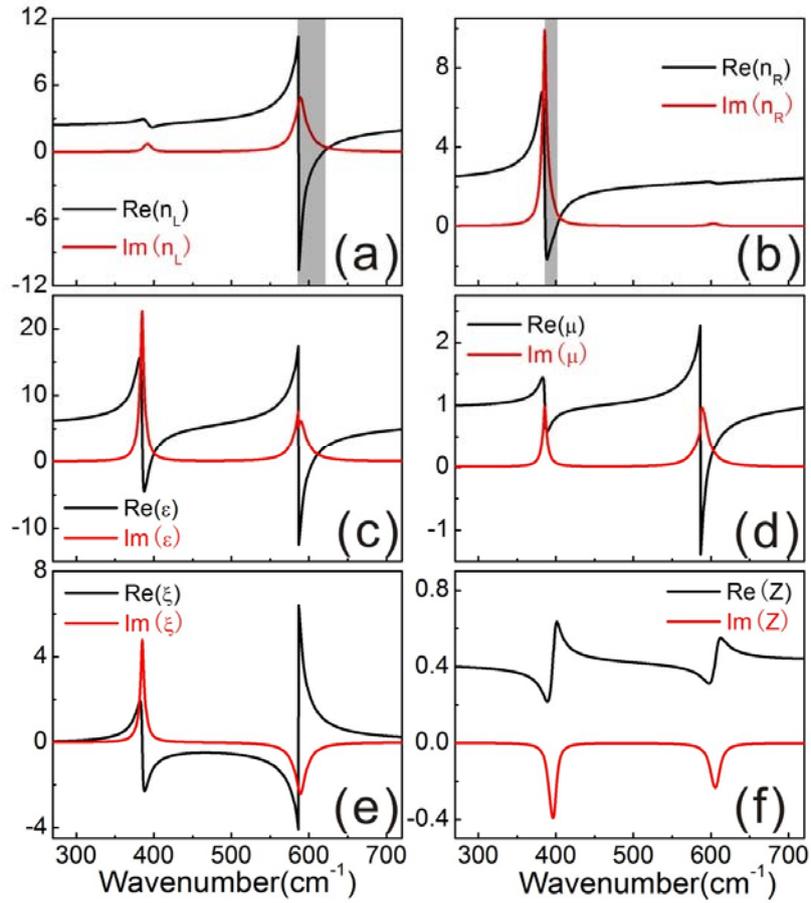

**Fig. 3** The retrieved effective optical parameters of USR arrays. **(a)** and **(b)** illustrate the real and imaginary parts of the refractive index for LCP and RCP light, respectively. The shadow denotes the occurance of negative refractive index. **(c)** and **(f)** illustrate the real and imaginary parts of the permittivity $\varepsilon$, permeability $\mu$, chiral parameter $\xi$ and the impedance Z.

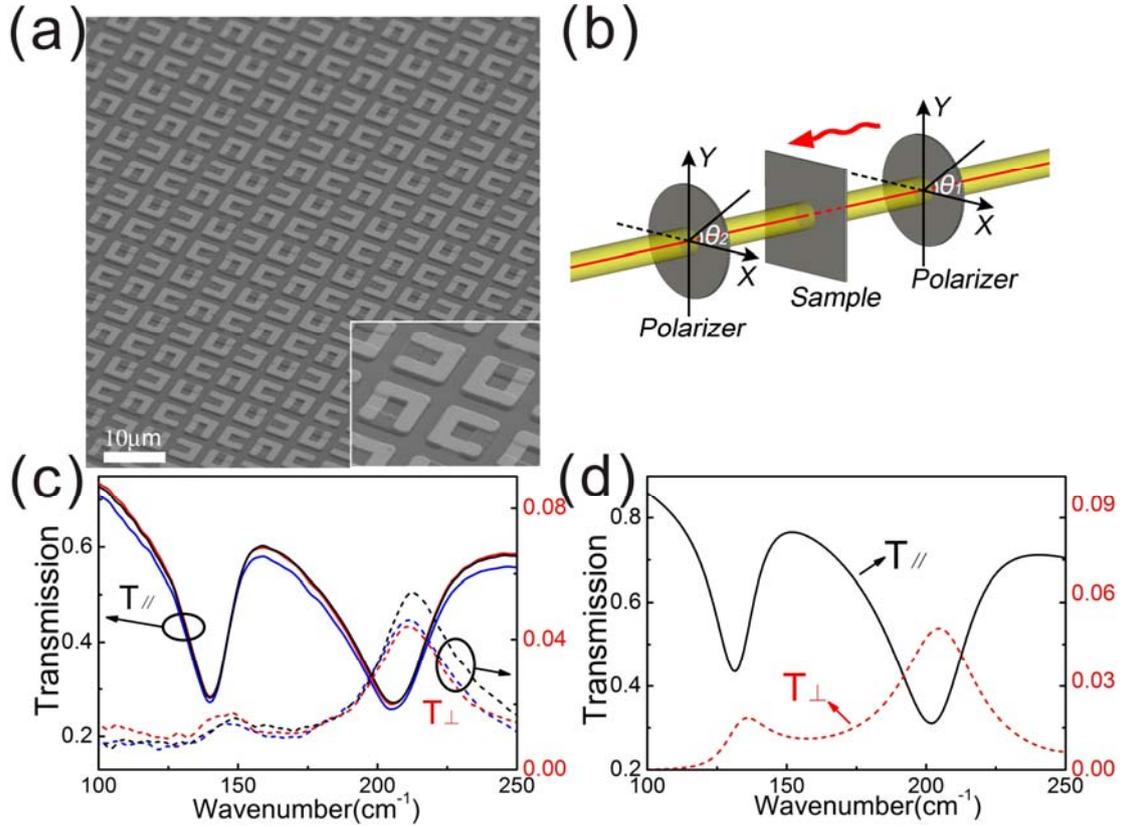

**Fig. 4 (a)** The scanning electron micrograph of the double-layered USR arrays. The double layer structure can be easily identified from the inset. **(b)** The schematics of measurement setup. In the measurement, the arms of USRs are set in parallel with X- and Y- directions. The polarizer in front of and after the sample can be independently rotated with angle $\theta_1$ and $\theta_2$, respectively. **(c)** Experimentally measured transmission spectra with different polarization of $\theta_1$ and $\theta_2$. The red solid line: $\theta_1=\theta_2= -45°$; the blue solid line: $\theta_1=\theta_2=0°$; the black solid line: $\theta_1=\theta_2=45°$; the red dash line: $\theta_1= -45$, $\theta_2=45°$; the black dash line: $\theta_1= 45, \theta_2= -45°$; the blue dash line: $\theta_1= 90°, \theta_2=0°$. **(d)** Simulation of the transmission $T_{//}$ and $T_\perp$. In simulation damping constant $\omega_\tau$ used in Drude model is doubled to fit the loss in real system.